\input{aipcheck}

\newcommand{\msun}{\mbox{$M_{\odot}$}}
\newcommand{\be}{\begin{equation}}
\newcommand{\ee}{\end{equation}}

\documentclass[
    ,final            
  ]
  {aipproc}

\layoutstyle{6x9}

\begin{document}

\title{SN Shock Evolution in the Circumstellar Medium surrounding SN 1987A}


\author{Vikram V. Dwarkadas}{
  address={Astronomy and Astrophysics, Univ.~of Chicago, 5640 S Ellis
Ave, AAC 010c, Chicago IL 60637} }

\begin{abstract}
 We study the structure of the circumstellar medium surrounding SN
 1987A in the equatorial plane. Furthermore, we study the evolution of
 the SN shock within this medium during the first 25 years, and the
 resulting hard X-ray and radio emission from the remnant.
 \end{abstract}

\maketitle


\section{INTRODUCTION}

The emission signatures from supernova (SN) 1987A showed significant
departures from those of other SNe. Radio emission was seen for a
period of a few months, died away for almost 2.5 years, was
re-observed around day 1150, and has since been increasing almost
linearly with time \cite{mgw02}. The soft X-rays show a more or less
linear increase from day 1100 till about day 6000, after which they
increase much more rapidly. The hard X-ray emission tends to agree
more with the radio \cite{pzb05} rather than the soft X-rays. The most
striking aspect of the optical emission is the gradually increasing
presence of "hot spots" along the inner ring beginning in 1997. The
most recent images (this conference) show that the spots are now
visible over almost the entire ring.

Observations have shown the presence of a bipolar nebula surrounding
the remnant, a probable result of the interaction between the blue
supergiant wind of the SN progenitor star with the red supergiant wind
from a previous epoch. The evolution of the radio and X-ray emission
led \cite[][hereafter CD95]{cd95} to propose the existence of an HII
region interior to this nebula. The properties of this HII region were
explored theoretically by \cite{pl99}, while observational
confirmation came via the work of Michael et al.~\cite{mmb98}.

Herein we reassess the properties of the circumstellar region around
SN 1987A in and around the equatorial plane, in light of recent radio
and X-ray observations. We study the evolution of the shock wave
within this region using numerical hydrodynamic simulations, compute
the hard X-ray and radio emission and compare to the observations.

\section{The Circumstellar Medium}

\begin{figure}
\label{fig:profile}
  \includegraphics[height=.4\textheight]{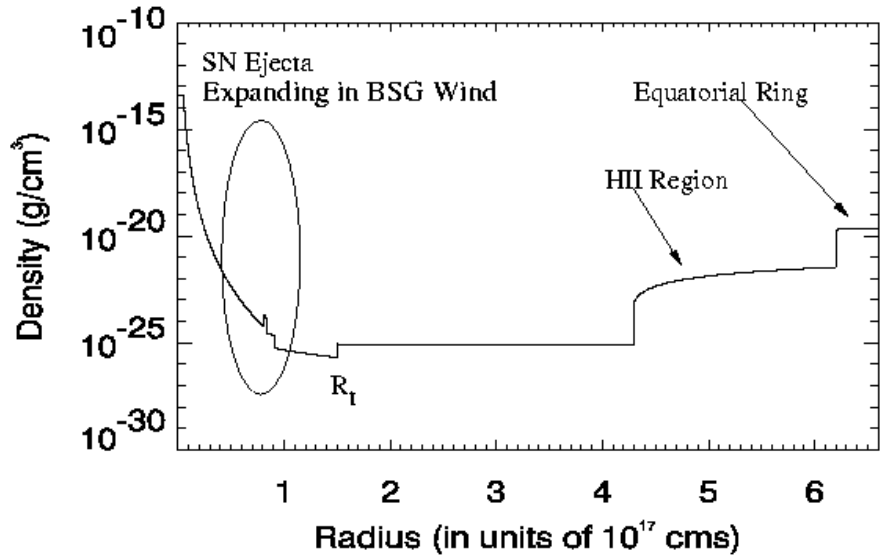} \caption{The
  initial SN ejecta and CSM density profile for the simulations.}
\end{figure}

Our intention is to model the cicumstellar medium (CSM) in a
spherically symmetric approximation, consistent with the observations
and theoretical ideas. The interaction of the blue supergiant (BSG)
wind with the red supergiant (RSG) wind gives rise to a wind-blown
bubble \cite{wmc77}. In the equatorial region, we find in order of
increasing radius (Fig.~\ref{fig:profile}): a freely expanding wind, a
wind-termination shock (R$_t$), a region of shocked wind, a region of
ionized wind (the HII region), and the dense shell or equatorial
ring. Based on the available information, CD95 had proposed values for
the radius of the wind termination shock, inner radius of the HII
region, velocity and mass-loss rate of the BSG wind, and the density
of the HII region. After more than a decade of further observations
providing new results on the radius and velocity of the SN shock wave,
and the X-ray and radio emission, we update this model to conform to
the latest observational data.

CD95 envisioned an HII region stretching from a radius r$_{\rm II}$ =
3 $\times 10^{17}$ cm until the inner edge of the equatorial ring
($\sim 6 \times 10^{17}$) cm. However, \cite{pl99} suggests that
r$_{\rm II} \sim 4.5 \times 10^{17}$ cm. Another limit comes from
\cite[][hereafter BG07]{bg07}. They find that the radio shell was about
0.3 pc across on day 1800. The radio emission was re-detected at about
1200 days, but the turn-on could have been a bit earlier. BG07 further
suggest that the initial velocity within the HII region was about 3600
km s$^{-1}$. Extrapolating backwards we find the inner edge of the HII
region to be around 4.3 $\times 10^{17}$ cm.

In this model the renewed X-ray and radio emission is due to the
interaction of the shock with the HII region. The detection of this
emission around day 1150 sets limits on the density of the medium in
which the shock is traveling, which is the freely-flowing BSG wind,
followed by the shocked BSG wind. The density increase of a factor of
4 beyond the wind-termination shock is too small to substantially
affect the dynamics of the SN shock, therefore the location of this
shock is unimportant for the dynamics. Following \cite{pl99} we take
the wind termination shock to be at a radius 1.5 $\times 10^{17}$.

In order for the SN shock to reach the HII region by day 1150, the
density of the freely expanding wind must be quite low. Assuming a
wind velocity of 550 km s$^{-1}$, consistent with that of CD95 and
other authors, the wind mass-loss rate must be about 5
$\times~10^{-9}~\msun$ yr$^{-1}$. We note that this mass-loss rate
seems quite low compared to known mass-loss rates of BSG
stars. However the density of the wind is tightly constrained, and the
wind velocity probably is well known within about 20-30\%. Then,
unless our estimate of the HII region radius is significantly off
(unlikely, given the observational data) the mass-loss rate must be
very low to account for the low wind density. A low mass-loss rate ($<
1.5-3 \times 10^{-7}~\msun$ yr$^{-1}$) was also postulated by
\cite{bl93} and \cite{lrc02}. The low density CS wind further reinforces the
suggestion (CD95) that synchrotron self-absorption, and not free-free
absorption, was responsible for the early radio absorption.

The density of the HII region is not well constrained, although
\cite{pl99} suggests that it is lower than that used by
CD95. Constraints arise from the radius and velocity evolution of the
shock wave, and the almost linearly increasing X-ray and radio
emission. We find that a density profile that increases gradually with
radius provides good results. The initial profile that we have adopted
for our simulations is shown in Figure 1.

\section{Numerical Simulations} The SN ejecta density is assumed to vary as r$^{-9}$, 
as suggested by \cite{lms94}. We take the ejecta profile to be
$\rho_{SN} = 9.3 \times 10^{37}~{\rm t}_{yr}^6~ {\rm r}_{pc}^{-9}$
following CD95.  Assuming spherical symmetry, we have carried out 1D
simulations to study the expansion of the SN ejecta into the
circumstellar medium (CSM) in the equatorial plane as described
above. The simulations are carried out using the VH-1 code, a
finite-difference hydrodynamics code. The parameters of the CSM were
adjusted based on the simulation to correctly reflect the observed
radius and velocity evolution of the SN forward shock as computed from
the radio \cite{bg07} and X-ray \cite{pzb05} data. Having computed the
evolution, we then calculate the hard X-ray and radio emission from
the remnant and compare to the observations.

\begin{figure}
\label{fig:radvel}
  \includegraphics[height=.3\textheight]{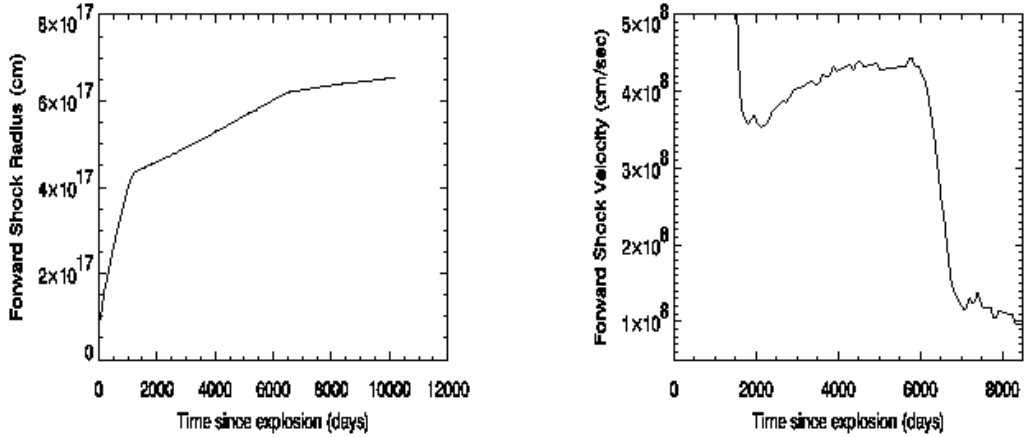} \caption{The
  radius (left) and velocity (right) evolution of the forward shock
  with time.}
\end{figure}

The interaction of the SN ejecta with the freely expanding wind leads
to the formation of a forward and reverse shock structure separated by
a contact discontinuity (see Fig.~\ref{fig:profile}). The forward
shock collision with the wind termination shock leads to a transmitted
shock expanding out into the shocked wind, and a reflected shock that
moves back into the ejecta \cite{vvd05}, overtaking the reverse
shock. The transmitted shock reaches the HII region at about day
1150. The impact slows down the shock considerably, and the velocity
of the shock transmitted into the HII region drops from $ >
30,000~{\rm km}~{\rm s}^{-1}$ to about 3500 km s$^{-1}$ (see
Fig.~\ref{fig:radvel}). The shock-HII region interaction leads to
renewed X-ray and radio emission (Fig.~\ref{fig:emiss}). Although the
shock is slowly sweeping up dense material from the HII region, its
velocity increases rather than decelerating
(Fig.~\ref{fig:radvel}). The reason is as follows: The impact with the
HII region results in an abrupt slowing down of the shock from its
velocity in the shocked wind region. Once the shock begins to expand
into the HII region it will approach another value with velocity much
less than what it was in the shocked region, but larger than the
velocity on HII region impact. The shock velocity increases after
impact until it transitions to this new value.

The transmitted shock collides with the inner edge of the equatorial
ring at about 18 years. This leads to a further slowing down of the
shock wave (Fig.~\ref{fig:radvel}), which loses a significant amount
of energy, gradually becoming radiative. The slowing down is reflected
in the radio and X-ray data, which indicate that the size of the radio
remnant is comparable to that of the inner edge of the equatorial
ring. The simulations show the presence of several reflected shocks
and rarefaction waves upon shock-ring impact.

\section{X-Ray and Radio Emission} 
\begin{figure}
\label{fig:emiss}
  \includegraphics[height=.3\textheight]{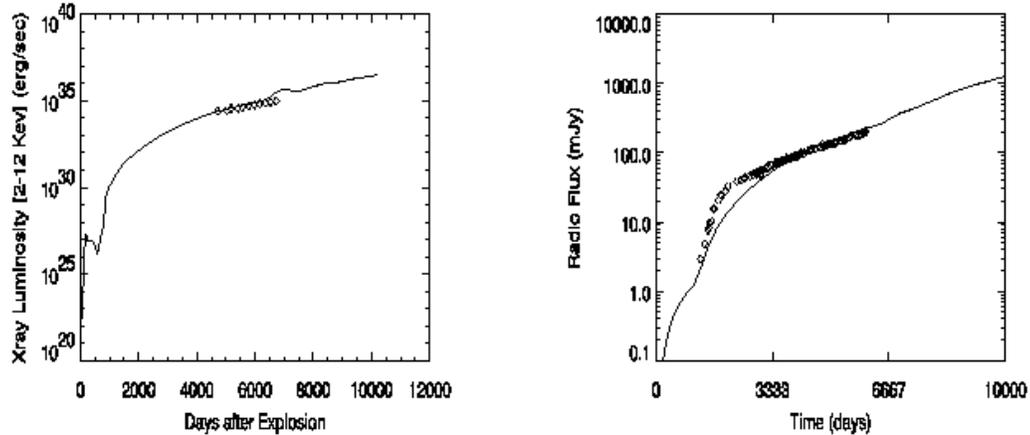} \caption{The
  evolution of the hard X-ray (left) and radio (right) emission from
  SN 1987A with time. The solid lines in each case is the calculation
  from our hydrodynamic model, the symbols refer to the data
  points. X-ray data is taken from \cite{pzb05} and radio data from
  \cite{mgw02}. }
\end{figure}

We compute the hard X-ray emission 
from our simulations in the range of 1-6 Angstroms, using the CHIANTI
database \cite{dlm97, l06}. This emission is presumably arising from
the shock-CSM interaction, whereas the soft X-ray emission likely
arises from the interaction of the shock wave with the finger-like
projections that give rise to the hot spots on the ring. Our results
compare well with the data presented in
\cite{pzb05}. Fig.~\ref{fig:emiss} shows the evolution of the X-ray
and radio emission with time.  The {\it hard X-ray emission arises
mainly from the reverse-shocked ejecta}, behind which the density is
low and temperature high. To match the magnitude of the emission
correctly we find that the electron temperature must be about 0.03
times the post-shock temperature, consistent with the theory presented
by \cite{glr06}. The forward shock velocity results in a low
post-shock temperature, thus its contribution to the hard X-ray
emission is minimal.

The radio emission was computed using the Chevalier mini-shell model
\cite{c82}. We find it difficult to reproduce the approximately
linear observed radio light curves \cite{mgw02}. Our best fit model
(Fig.~\ref{fig:emiss}) is unusual in that it requires either the
magnetic field or the relative particle energy density to be constant
with time (see \cite{vvd06}). Figure~\ref{fig:emiss} shows that our
model also is unable to accurately reproduce the rapid rise seen after
day 1150. However we note that the impact with the HII region will
result in a reflected shock which can also accelerate particles,
likely enhancing the radio emission for a short period \cite{c92}.

There are undoubtedly shortcomings to a spherically symmetric approach
in this case. It is therefore heartening to see that such a simplistic
approach reasonably reproduces the shock dynamics and kinematics, and
the increasing behavior of the X-ray and radio flux with time. It also
lays down a platform from which to proceed to do more complicated
multi-dimensional simulations in future, that take the asymmetries of
the structure, and the presence of hydrodynamical instabilities, into
account.

\begin{theacknowledgments}
VVD is supported by award \# AST-0319261 from the National Science
Foundation, and by NASA through grant \# HST-AR-10649 awarded by
STScI. I would like to acknowledge several extremely helpful
discussions with R.~Chevalier, B.~Gaensler and S.~Park. The organizers
deserve thanks for a well-organized workshop in a beautiful location,
and Kurt Weiler for making sure that all sessions ran on time.
\end{theacknowledgments}





\begin{thebibliography}{99}

\bibitem{bl93} Blondin, J. M., \& Lundqvist, P.,
\emph{ApJ},\textbf{405}, 337--352, (1993)

\bibitem{dlm97} Dere, K.~P., et al., \emph{A\&AS},  \textbf{125}, 149--173 (1997)

\bibitem{cd95}
Chevalier, R.~A., \& Dwarkadas, V.~V., \emph{ApJ}, \textbf{452},
L45--L48, (1995)


\bibitem{c92}
Chevalier, R. A.~\emph{Nature}, \textbf{355}, 617--618, (1992)

\bibitem{c82}
Chevalier, R. A.~\emph{ApJ}, \textbf{259}, 302--310, (1982)

\bibitem{vvd06}
Dwarkadas, V.~V., ``Supernova Explosions in Winds and Bubbles, with
Applications to SN 1987A'', To appear in RMxAA, 2007
(astro-ph/0612665)

\bibitem{vvd05} Dwarkadas, V.~V., \emph{ApJ}, \textbf{630}, 892--910, (2005)

\bibitem{bg07}
Gaensler, B., et al.~in "Supernova 1987A: 20 Years After: Supernovae
and Gamma-Ray Bursters" AIP, New York, eds. S. Immler, K.W. Weiler,
and R. McCray, (2007)

\bibitem{glr06} Ghavamian, P., Laming, J.~M., \& Rakowski, C.~E.,
ApJL, accepted (astroph/0611306) (2006)

\bibitem{l06} Landi, E.,~et al., \emph{ApJS}, \textbf{162}, 261--280, (2006)

\bibitem{lrc02} 
Link, R., et al.~ \emph{Interacting Winds from Massive Stars}, eds.~
A.~F.~J.~Moffat \& N.~St.-Louis, \textbf{ASP Conf.~Series 260}, ASP,
New York, (2002)

\bibitem{pl99}
Lundqvist, P., \emph{ApJ}, \textbf{511}, 389--397, (1999)

\bibitem{lms94} Luo, D., McCray, R., \& Slavin, J., \emph{ApJ}, \textbf{430}, 264--276, (1994)

\bibitem{mgw02} 
Manchester, R.~N., Gaensler, B.~M., et al., \emph{PASA}, \textbf{19},
207--221, (2002)

\bibitem{mmb98} 
Michael, E., McCray, R., Borkowski, K.~J., Pun, C.~S.~J., Sonneborn,
G., \emph{ApJL}, \textbf{492}, 143--146, (1998)

\bibitem{pzb05}
Park, S., Zhekov, S.~A., Burrows, D.~N., \& McCray, R., \emph{ApJ},
\textbf{634}, 73--76, (2005)


\bibitem{wmc77} Weaver, R., McCray, R.,  Castor, J.,
Shapiro, P., \& Moore, R., \emph{ApJ}, \textbf{218}, 377--395, (1977)


\end{thebibliography}
\end{document}